\documentclass[12pt,draftclsnofoot,onecolumn]{IEEEtran}
\usepackage{MnSymbol}
\usepackage{lipsum}
\usepackage{optidef}
\usepackage{amsmath}
\usepackage{bm}
\usepackage{amsfonts}
\usepackage{paralist}
\usepackage{cite}
\usepackage{graphicx}

\makeatletter
\newcommand{\bol}[1]{\ensuremath{\boldsymbol{#1}}}
\makeatother
\begin{document}
\title{Secure Simultaneous Information and Power Transfer for Downlink Multi-user Massive MIMO}
\author{ \small Zahra Goli$^\text{1}$, S. Mohammad Razavizadeh$^\text{1, 3}$, Hamed Farhadi$^\text{2}$, Tommy Svensson$^\text{3}$ \ \\ \ \\
 {\footnotesize $^\text{1}$School of Electrical Engineering, Iran University of Science and Technology (IUST), Tehran, Iran \\
\footnotesize $^\text{2}$Ericsson, Stockholm, Sweden and KTH Royal Institute of Technology, Stockholm, Sweden \\ 
\footnotesize $^\text{3}$Chalmers University of Technology, Sweden \\
\footnotesize z{\_}goli@elec.iust.ac.ir, smrazavi@iust.ac.ir, farhadih@kth.se, tommy.svensson@chalmers.se}}
 
\maketitle
 
 \normalsize

\begin{abstract}
	In this paper, downlink secure transmission in simultaneous information and power transfer (SWIPT) system enabled with massive multiple-input multiple-output (MIMO) is studied. A base station (BS) with a large number of antennas transmits energy and information signals to its intended users, but these signals are also received by an active eavesdropper. The users and eavesdropper employ a power splitting technique to simultaneously decode information and harvest energy. Massive MIMO helps the BS to focus energy to the users and prevent information leakage to the eavesdropper. The harvested energy by each user is employed for decoding information and transmitting uplink pilot signals for channel estimation. It is assumed that the active eavesdropper also harvests energy in the downlink and then contributes during the uplink training phase. Achievable secrecy rate is considered as the performance criterion and a closed-form lower bound for it is derived. To provide secure transmission, the achievable secrecy rate is then maximized through an optimization problem with constraints on the minimum harvested energy by the user and the maximum harvested energy by the eavesdropper. Numerical results show the effectiveness of using massive MIMO in providing physical layer security in SWIPT systems and also show that our closed-form expressions for the secrecy rate are accurate.
\end{abstract}
\section{Introduction}
Recently, employing energy harvesting techniques has been regarded as a promising approach to prolong the lifetime of wireless low power networks. These techniques are useful in many applications including wireless communications in extreme environments, sensor networks, and medical Internet of Things (m-IoT) applications \cite{RFmagazin,RFsensors,farhadi2019medical}. Traditional renewable energy sources such as solar and wind energy are weather dependent and not available everywhere and anytime. Thus, energy harvesting from ambient radio frequency (RF) signals has recently drawn a significant research interest due to many practical advantages, such as wide operating range, being predictable, low production cost, small receiver form factor, and efficient energy multicasting thanks to the broadcast nature of electromagnetic (EM) waves\cite{RFmagazin,RFsensors,farhadi2019medical,sensitivenonlinear,RFsurvey,ding2015application}. The wide coverage of cellular communication networks, and the need for powering a massive number of low-power IoT devices in the next generations of wireless networks provide an opportunity for RF-based power transfer techniques to be considered as a prominent and scalable solution.
The conventional role of RF signals as information carrier has attracted attention to simultaneous wireless information and power transfer (SWIPT) as an emerging technology to solve the energy supply problem in power- constrained networks. In SWIPT systems, two architectures of \textit{separated} and \textit{hybrid} receivers have been proposed for information decoding and energy harvesting. In the separated architecture, the information and energy receivers operate separately \cite{SWIPT2013,SWIPTMISO,phasenoise}, whereas in the hybrid architecture a common receiver is used for both harvesting energy and information decoding. In this architecture, the signal which is used for decoding the information cannot be reused for harvesting the energy due to hardware limitations\cite{zhou2013wirelessArchitecture}. Thus, the received signal has to be split into two parts, one for information decoding and another for energy harvesting. Time switching and power splitting are two common hybrid receiver architectures in the literature\cite{SWIPTM,jointTSPS,timeswitchingSWIPT,powersplitting}. On the other hand, in SWIPT systems, the RF energy harvesting in the receiver is generally modeled by two linear and non-linear models  which the non-linear model is more practical \cite{massiveSWIPT2018,massiveSWIPT2017,sensitivenonlinear,nonlinearRectifier, schobernonlinear1,schobernonlinear2,energynonelinear,energynonelinear2}. 

Since RF signals significantly attenuate over distance, improving energy transfer efficiency is a great challenge for deploying SWIPT systems over wide areas and especially in applications like IoT. To improve the efficiency of energy transfer in SWIPT systems, various techniques can be adopted. In \cite{rostampoor2017energy,zhao2015energy}, the authors propose methods based on using relay techniques to achieve this goal. Another solution for improving the efficiency of energy transfer is using massive multiple-input multiple-output (MIMO). By employing a large number of antennas, massive MIMO can provide extremely narrow beams towards desired users to efficiently transfer energy to them. Even though massive MIMO can play a significant role in SWIPT systems, the research in this area is in its infancy. SWIPT enabled massive MIMO systems have been investigated in \cite{jointTSPS,massiveSWIPT2017,massiveSWIPT2018,massiveMIMOSWIPTtilt}. In \cite{jointTSPS}, a hybrid time switching and power splitting SWIPT protocol design in a full-duplex massive MIMO system was proposed and the achievable rate was maximized by optimizing transmit powers of the base station (BS). SWIPT for downlink of a multi-user massive MIMO system was studied in \cite{massiveSWIPT2017} and its achievable rates were computed. In \cite{massiveSWIPT2018}, a massive MIMO SWIPT system assuming Rician fading channels was investigated and the approximate achievable rate and harvested energy were derived. In \cite{massiveMIMOSWIPTtilt}, SWIPT in a 3D massive MIMO system was studied and the BS antenna's tilt was optimized jointly with power allocation and power splitting ratios in order to increase SWIPT efficiency. \\
Due to the broadcast nature of the wireless channels, wireless networks are always vulnerable to physical layer attacks including eavesdropping and jamming. This problem is more challenging in SWIPT systems, since the transmitted energy can help the attackers. Traditionally, communication security relies on cryptography techniques. Encryption and decryption algorithms are usually complex and energy consuming. Therefore, these techniques are not suitable to provide security for use in energy limited networks and SWIPT systems. Physical layer security has recently received significant research interests to guarantee secure communication in SWIPT systems by utilizing physical properties of wireless channels such as channel fading, noise and interference \cite{ArtificialNoise2015,robustSWIPTsecure,beamformingartificialnoise,beamformingartificialnoisecognitive,securebeamforming}. In \cite{ArtificialNoise2015}, a secure transmission scheme was proposed by exploiting artificial noise. The authors in \cite{robustSWIPTsecure} exploit the energy signal in addition to artificial noise to confound eavesdropper and provide secure communication. In \cite{beamformingartificialnoise,secureSWIPTmagazin,beamformingartificialnoisecognitive,securebeamforming}, security of communication in SWIPT systems was provided by exploiting optimal beamforming design.  Recently massive MIMO has attracted attention to ensure security of transmission. Due to the high spatial resolution provided in massive MIMO, information leakage to illegal receivers can be reduced \cite{physicallayermassiveMIMO,razavi,securemassiveMIMOmulticell,hybridsecuremassiveMIMO,pirzadeh2016subverting}. However, a so called pilot contamination attack in the training phase can help an active eavesdropper to wiretap the signals in massive MIMO systems \cite{pirzadeh2016subverting,akhlaghpasand2017jamming} 

In SWIPT systems, massive MIMO can combat eavesdropping in addition to improve energy and information transfer efficiency. In spite of these advantages, research on this topic has received scant attention and more research is necessary. In \cite{phasenoise}, a SWIPT system with massive MIMO and separate energy and information receivers was considered. The effect of phase noise on the accuracy of channel state information estimation and information leakage to energy receivers that are potential eavesdroppers, was studied in \cite{phasenoise}. In \cite{SWIPTactiveeavesdropper}, a multi-cell massive MIMO system in the presence of a two-antenna active energy harvester was studied. The energy harvester legitimately harvests energy via one antenna and illegitimately eavesdrops the signal transmitted for information users via the other antenna. The power allocation for downlink transmission is optimized according to the asymptotic lower bound on averaged harvested energy and ergodic secrecy rate. \newline
Motivated by the above works, in this paper, we study secure transmission in the downlink of a multi-user massive MIMO SWIPT system. A BS is equipped with a large number of antennas and simultaneously transmits energy and data signals to its intended users. In addition, an active eavesdropper in the area wiretaps the signals transmitted by the BS. Each transmission phase is composed of an uplink training (pilot transmission) phase and a downlink energy and data signals transmission phase.  
In the uplink training phase, the users send their allocated pilots to the BS and the BS uses them to estimate the users' channels. 
We assume the system operates in the time division duplex (TDD) mode and because of the channel reciprocity property, the estimated uplink channels are then used by the BS for downlink beamforming design. To facilitate eavesdropping, the eavesdropper contaminates the uplink training phase and sends a pilot signal to the BS simultaneously with the users. In this network, both users and the active eavesdropper employ the hybrid receivers and decode information and harvest energy by using a power splitting method.
In the above network, we analyze the achievable secrecy rate and derive an accurate closed-form lower bound for it. Then, using the derived lower bound, the achievable secrecy rate is maximized by proper choice of the
power splitting ratio and the fraction of the harvested energy allocated to the uplink pilot training. In the resulting optimization problem, we consider constraints on the minimum harvested energy by the user and the maximum harvested energy by the eavesdropper. Numerical results show that using massive MIMO can significantly enhance the security performance of the SWIPT networks. We also show that the derived lower bound is very close to the actual achievable secrecy rate.   \newline
\textit{Notation}: Boldface lowercase and uppercase letters denote vectors and matrices, respectively. The superscript $(.)^∗$, $(.)^T$ and $(.)^H$  represent conjugate, transpose and conjugate
transpose, respectively. $(.)_{\text{re}}$ and $(.)_{\text{im}}$ denote real part and imaginary part respectively. The notations $\lvert.\rvert$ and $\lVert.\rVert$ represent absolute value and 2-norm. $\mathbb{E}\{.\}$ and $var(.)$ stand
for expectation and variance operations. $\mathcal{C}\mathcal{N}(\mu,\sigma^2)$ denotes the circular symmetric complex Gaussian
distribution with mean $\mu$ and covariance  $\sigma^2$. The notation $[x]^+$ denotes $max(x, 0)$.
\section{System Model}
\begin{figure}
	%\hspace{0.7cm}
	\centering
	\includegraphics[scale=0.32]{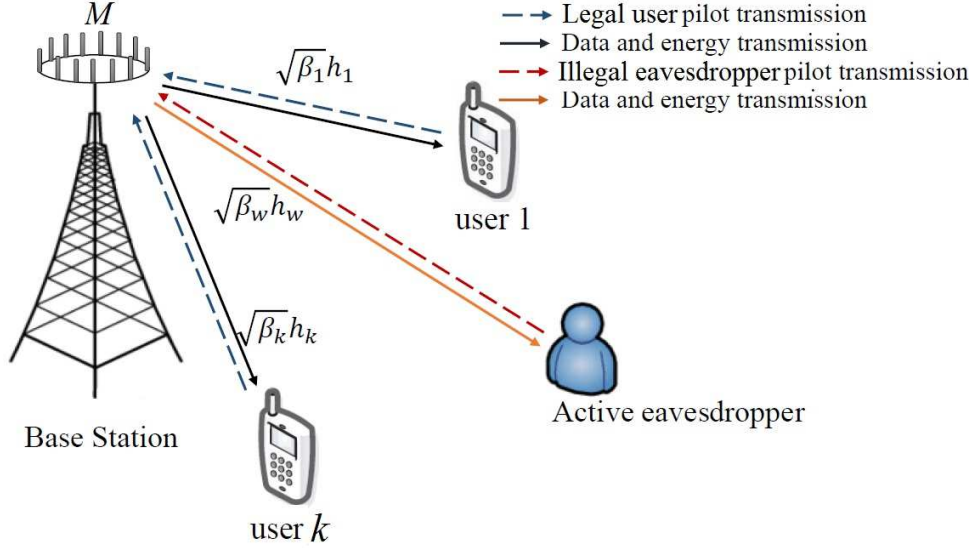}
	\caption{System model of a SWIPT system consisting of a BS equipped with \textit{M} antennas, \textit{K} single antenna users and a single antenna active eavesdropper. The BS has a continuous and stable power supply, while all the users and the eavesdropper operate with the wireless EH.}
	\label{fig:system}
\end{figure}

A massive MIMO system consisting of a BS equipped with \textit{M} antennas, \textit{K} single antenna users and an active eavesdropper as shown in Fig. \ref{fig:system} is considered. The \text{BS} transmits data to the users who are able to simultaneously decode the information and harvest energy from the radio signals. Also, the eavesdropper illegally receives the transmitted signal and decodes information and harvests energy. It is assumed that the BS, users and eavesdropper are synchronized and operate using the TDD protocol. A frame-based transmission in one coherence interval \textit{T}
consisting of two phases %*as shown in Fig. \ref{fig:Frame}%*
is considered. In the first phase with length $\tau$ (i.e. the training phase), the users simultaneously transmit orthogonal pilots to the BS. Also, the eavesdropper which aims to receive users information illegally, simultaneously transmits a pilot sequences to disturb the channel estimation. Since the eavesdropper does not know the users' pilot sequence, it chooses a random pilot sequence uniformly distributed over the unit sphere \cite{pirzadeh2016subverting}.
\\ After receiving the signals in this phase by the BS, uplink channels are estimated by the BS. The BS then determines the downlink channel by assuming the channel reciprocity. In the next phase, which is dedicated to data and energy transmission, the BS transmits information and energy signals to the users using maximal ratio transmission (MRT), and users harvest the RF energy from the BS and simultaneously decode information based on the following power-splitting technique. In the assumed power splitting technique, the received signal is split into two power streams with power splitting ratio $1-\rho_k$ and $\rho_k$ for harvesting energy and decoding information, respectively ($0<\rho_k<1$). The hybrid receiver is a generalization of a conventional information receiver and an energy harvesting receiver. In particular, by setting the power splitting ratios as $\rho_k=1$ and $\rho_k=0$, the hybrid receiver reduces to an information receiver and an energy harvesting
receiver, respectively. 

Let $ \bm{g}_{k}=\sqrt{\beta_{k}}\bm{h}_{k} $  denote the channel vector between the BS and the $k$th user, where $ \beta_{k} $ and $ \bm{h}_{k} \thicksim \mathcal{C}\mathcal{N}(0,\bm{I}_M) $ represent the large scale fading and small scale fading of the channel, respectively. Furthermore, $\bm{g}_{w}=\sqrt{\beta_{w}}\bm{h}_{w}$ is the channel vector between eavesdropper and BS where $\beta_w$ and $\bm{h}_{w}\thicksim \mathcal{C}\mathcal{N}(0,\bm{I}_M)$ are the large scale fading and small scale fading of the channel, respectively. 

%\begin{figure}[!h]
%	\centering
%	\includegraphics[scale=0.57]{Frame.png}
%	\caption{Transmission Frame. Each transmission frame consists two phases. The users legally and the active eavesdropper illegally transmit pilot to the BS in the first $\tau$ time using the energy harvested in the previous
%		frame. After the training phase, the BS transmits data and energy signal
%		simultaneously to the users in the remaining $T-\tau$ time. }
%	\label{fig:Frame}
%\end{figure}
The channel between all users and the BS can be represented in matrix form as 
\begin{align} \label{channel matrix}
\textbf{G}=\textbf{H} \textbf{D}^{\frac{1}{2}},
\end{align}
where $\textbf{H}=\left[ \bm{h}_{1},...,\bm{h}_{k},...,\bm{h}_{K}\right] $ and \textbf{D} is a diagonal matrix whose  elements  $\left[ \textbf{D}\right] _{kk}=\beta_{k}$.
\subsection{Training Phase}
The pilot sequences used by the users can be represented by an $ \eta\times K $ matrix $\sqrt{\eta p_{t}}\Phi$, where $ \bm{\phi}_{k} $, the $k$th user pilot sequence is the $k$th column of $ \Phi $, $\eta$ represents the pilot sequence length and $\Phi^{H}\Phi=\textbf{I}_{k} $. The received signal at the BS is\\
\begin{equation} \label{received signal at BS}
\bol{Y}_{t}=\sqrt{\eta p_{t}}\bol{G}\Phi^{T}+\sqrt{\eta q_{t}}\bol{g}_{w}\bol{\phi}_{w}^{T}+\textit{\bol{N}},  
\end{equation}
where $\bm{ \phi}_{w} $ is active the eavesdropper's pilot sequence and $\textit{\bol{N}}$ is an $ M\times \eta $ noise matrix with i.i.d $ \mathcal{C}\mathcal{N} (0,\sigma^2) $ elements. The energy allocated to each pilot symbol by the users and eavesdropper are denoted by $p_t$ and $q_t$, respectively and defined as
\begin{align}
 & p_t=\dfrac{\theta Q^{NL}_k}{\eta} \\ 
 &q_t=\dfrac{\zeta Q^{NL}_{\text{Eve}}}{\eta}, 
\end{align} 
where $Q^{NL}_k$ and $Q^{NL}_{\text{Eve}}$ are total harvested energy by the $k$th user and the eavesdropper, respectively. The parameters  $\theta \in[0,1]$ and $\zeta \in [0,1]$ denote the fraction of total harvested energy allocated to the pilot phase by the users and the eavesdropper, respectively.\\
The minimum mean squared error (MMSE) estimate of \textbf{G} given $\textbf{Y}_{t} $ is \cite{kay1993fundamentals}
\begin{equation} \label{MMSE}
\hat{\bm{G}}=\dfrac{1}{\sqrt{\eta p_{t}}}\bm{Y}_{t}\Phi^{*}\left(\bm{I}_{k}+\frac{\sigma^2+q_{t}\beta_{w}}{\eta p_{t}}\bm{D}^{-1}\right)^{-1}.
\end{equation}
Define $\bm{\mathcal{E}} \triangleq \hat{\bm{G}}-\bm{G}$. Then we have
\begin{equation}
\sigma_{\hat{g}_k}^{2}=\dfrac{\eta p_t \beta_k  ^{2}}{\sigma^2+q_t\beta_w +\eta p_t \beta_k}  
\end{equation}
\begin{equation}
\sigma_{e_k}^{2}=\frac{\left(\sigma^2+q_t\beta_w\right)\beta_k}{\sigma^2+q_t\beta_w +\eta p_t \beta_k},
\end{equation}\\
where $\sigma_{\hat{g}_k}^{2}$ and $\sigma_{e_k}^{2}$ are the variances of the independent zero mean
elements in the $k$th column of $\hat{\bm{G}}$ and $\bm{\mathcal{E}}$, respectively.
\subsection{Data and Energy Transmission Phase}
In this phase, BS transmits information and energy signals to all users using MRT precoding which is optimal precoding in the massive MIMO regime\cite{bjornson2017massive}. The energy signal is a pseudorandom signal which is perfectly known at the transmitter and the users. It does not carry any information and it is transmitted by the BS not only to harvest more energy by the users but also to confound the eavesdropper. An eavesdropper that interfered in the training phase and succeeded in redirecting the signal beam toward itself, can harvest energy and decode information illegally in this phase. The received signal by the $k$th user and eavesdropper can be expressed respectively as
\begin{align}
&y_k  =\bm{g}_k^H \sum_{i=1}^{K}\bm{w}_i \left(s_i+w_{E_i}\right)+n_{\text{ant,k}} \\
&y_{Eve}  =\bm{g}_w^H \sum_{i=1}^{K}\bm{w}_i \left(s_i+w_{E_i}\right)+n_{\text{ant,Eve}}
\end{align}
where $s_k\in\mathbb{C}$ and $w_{E_k}\in\mathbb{C}$ denote the information and energy symbols for the $k$th user. Without loss of generality, it is assumed that $\mathbb{E}\{\lVert s_k\rVert^2\}=1$ and $\mathbb{E}\{s_k\}=0$. Also, $\mathbb{E}\{\lVert w_{E_k}\rVert^2\}=W_E$ and $\mathbb{E}\{w_{E_k}\}=0$. Furthermore, $\bm{w}_k=\dfrac{\hat{g}_k}{\sqrt{\mathbb{E}\{\lVert\hat{g}_k\rVert^2\}}}$ represents the MRT precoding vector of the $k$th user. Also, $n_{ant,k} \sim \mathcal{C}\mathcal{N} (0,\sigma_{ant}^2)$ and $n_{ant,Eve} \sim \mathcal{C}\mathcal{N} (0,\sigma_{ant}^2)$ denote additive white Gaussian noise (AWGN) at each receiver and eavesdropper, respectively.  \\
\section{Harvested Energy and Achievable Secrecy Rate}  
In this section, averaged harvested energy and achievable secrecy rate are analysed and lower bounds on achievable secrecy rate is derived. Achievable secrecy rate is a common criterion to assess security of transmission and defined as the rate difference between the main channel from the BS to the user and the wiretap channel from the BS to the eavesdropper.
%\newcounter{Mytempeqncnt}
%\begin{figure*}[!t]
%	\normalsize
%	\setcounter{Mytempeqncnt}{\value{equation}}
%	\setcounter{equation}{22}
%	\begin{align} \label{secrecylowerbound}
%	\text{R}_{\text{S,bound}}&=\dfrac{T-\tau}{T}\log_2\left(1+\dfrac{\rho_kM\eta p_t \beta_k^2/\left(\sigma^2+q_t \beta_w+\eta p_t \beta_k\right)}{\rho_k\left(K\beta_k+(k-1)\beta_kW_E+\dfrac{\beta_k(q_t \beta_w+\sigma^2)}{\eta p_t\beta_k+q_t \beta_w+\sigma^2}+\sigma_{ant}^2 \right)+\sigma_s^2}\right)\nonumber\\
%	&-\dfrac{T-\tau}{T}\log_2\left(1+\dfrac{\rho_{\text{Eve}}M\eta q_t \beta_w^2/\sigma^2+q_t \beta_w+\eta p_t \beta_k}{\rho_{\text{Eve}} \left(\left(K\beta_w+\sum_{i=1}^K \dfrac{Mq_t \beta_w^2}{\sigma^2+q_t \beta_w+\eta p_t \beta_i }\right)\left(W_E+1\right)-\dfrac{M q_t \beta_w^2}{\sigma^2+q_t \beta_w+\eta p_t \beta_k}+\sigma_{ant}^2\right)+\sigma_s^2}\right)
%	\end{align}	
%	\setcounter{equation}{\value{Mytempeqncnt}}
%	\hrulefill
%	%\vspace*{4pt}
%\end{figure*}
%A. Average Harvested Energy\\
\subsection{Average Harvested Energy}
In this section, the harvested energy and achievable secrecy rate are analyzed and lower bound on achievable secrecy rate is derived. Achievable secrecy rate is a common metric to assess security of transmission and defined as the rate difference between the main channel from the BS to the user and the wiretap channel from the BS to the eavesdropper. 
%\newcounter{Mytempeqncnt}
%\begin{figure*}[!t]
%	\normalsize
%	\setcounter{Mytempeqncnt}{\value{equation}}
%	\setcounter{equation}{20}
%	\begin{align} \label{secrecylowerbound}
%	\text{R}_{\text{S,bound}}&=\dfrac{T-\tau}{T}\log_2\left(1+\dfrac{\rho_kM\eta p_t \beta_k^2/\left(\sigma^2+q_t \beta_w+\eta p_t \beta_k\right)}{\rho_k\left(K\beta_k+(k-1)\beta_kW_E+\dfrac{\beta_k(q_t \beta_w+\sigma^2)}{\eta p_t\beta_k+q_t \beta_w+\sigma^2}+\sigma_{ant}^2 \right)+\sigma_s^2}\right)\nonumber\\
%	&-\dfrac{T-\tau}{T}\log_2\left(1+\dfrac{\rho_{\text{Eve}}M\eta q_t \beta_w^2/\sigma^2+q_t \beta_w+\eta p_t \beta_k}{\rho_{\text{Eve}} \left(\left(K\beta_w+\sum_{i=1}^K \dfrac{Mq_t \beta_w^2}{\sigma^2+q_t \beta_w+\eta p_t \beta_i }\right)\left(W_E+1\right)-\dfrac{M q_t \beta_w^2}{\sigma^2+q_t \beta_w+\eta p_t \beta_k}+\sigma_{ant}^2\right)+\sigma_s^2}\right)
%	\end{align}	
%	\setcounter{equation}{\value{Mytempeqncnt}}
%	\hrulefill
%	%\vspace*{4pt}
%\end{figure*}
\subsection{Average Harvested Energy} 
The received signal is split into two power streams with power splitting ratios $\rho_k$ and $1-\rho_k$ to respectively decode information and harvest energy by the $k$'th user\cite{massiveSWIPT2017}. Also, the received signal is split  into two power streams with power splitting ratios $\rho_{Eve}$ and $1-\rho_{Eve}$ by the eavesdropper. 
The harvested energy by receivers in SWIPT systems is modeled either as ideal linear \cite{massiveSWIPT2018,massiveSWIPT2017} or more realistically using non-linear models \cite{schobernonlinear1,schobernonlinear2}.
% In linear model\cite{massiveSWIPT2018,massiveSWIPT2017}, the harvested energy can be expressed as
%\begin{align}
%Q^L_{k}=\alpha(T-\tau)P^{EH}_{k}
%\end{align}
%Where $\alpha\in[0,1]$ represents conversion efficiency and $P^{EH}_{k}$ denotes the total received RF power at the users. 
%Similar to users, the linear harvested energy for eavesdropper can be written as
%\begin{align}
%Q^L_{Eve}=\alpha_{Eve}(T-\tau)P^{EH}_{Eve}
%\end{align}
%Where $\alpha_{\text{Eve}}\in[0,1]$ represents conversion efficiency and $P^{EH}_{Eve}$ denotes the total received RF power at %the eavesdropper.\\
In the non-linear model, the sensitivity of the energy harvester is limited \cite{energynonelinear}. Thus, the harvested energy is modeled using the logistic (or sigmoid)
function at the users and eavesdropper as follows\cite{energynonelinear,energynonelinear2}.

%\begin{shaded}
\begin{align}
Q^{NL}_{\text{k}}&=\dfrac{P_{s_k}}{\text{exp}(a\times b)}\times\left[ \dfrac{1+\text{exp}(a\times b)}{1+\text{exp}\left(-a([P^{EH}_{k}-P_{SEN}]^{+}-b)\right)}-1\right](T-\tau),
\end{align}
\begin{align}
Q^{NL}_{\text{Eve}}&=\dfrac{P_{s_{Eve}}}{\text{exp}(a\times b)}\times\left[ \dfrac{1+\text{exp}(a\times b)}{1+\text{exp}\left(-a([P^{EH}_{Eve}-P_{SEN}]^{+}-b)\right)}-1\right](T-\tau),
\end{align}
%\end{shaded}
where $P_{s_k}$ and $P_{s_{\text{Eve}}}$  denotes the maximum amount of harvested power when the EH circuit is saturated at the users and eavesdropper, respectively. Here, a and b are positive constants related to the circuit specification. $P_{SEN}$ denotes the EH sensitivity, that is, the harvester can only collect the RF energy when the input power is greater than $P_{SEN}$. It is assumed that all the users and eavesdropper apply the same non-linear energy harvester.
The harvested energy is used for information decoding and pilot transmission in the next frame by the users and the eavesdropper.
The $P^{EH}_{k}$ and $P^{EH}_{Eve}$ which are respectively the total harveted power at the $k$'th user and eavesdropper can be  expressed as
\begin{align} \label{Quser}
P^{EH}_{\text{k}}=&(1-\rho_k) \mathbb{E}\left\lbrace\left| \bm{g}_k^H \sum_{i=1}^{K}\bm{w}_i \left(s_i+w_{E_i}\right)+n_{ant,k}\right|^2\right\rbrace,  
\end{align}
\begin{align}
P^{EH}_{\text{Eve}}=&(1-\rho_{Eve})\mathbb{E}\left\lbrace\left| \bm{g}_w^H \sum_{i=1}^{K}\bm{w}_i \left(s_i+w_{E_i}\right)+n_{ant,Eve}\right|^2\right\rbrace \label{Q eaves}.
\end{align}
\\ \\
\textbf{Theorem 1:} By exploiting MRT precoding, the total received RF power by the $k$'th user and eavesdropper can be respectively obtained as
\begin{align} 
P^{EH}_{k}&=(1-\rho_{k})\Bigg(\left(K\beta_k+\dfrac{M\eta p_t \beta_k^2}{\sigma^2+q_t \beta_w+\eta p_t \beta_k}\right)\left(W_E+1\right)+\sigma_{ant}^2\Bigg),\label{calculated user energy}
\end{align}
\begin{align} \label{Q eaves2}
P^{EH}_{\text{Eve}}&=(1-\rho_{\text{Eve}}) \Bigg(\left(K\beta_w+\sum_{i=1}^K\dfrac{M q_t \beta_w^2}{\sigma^2+q_t \beta_w+\eta p_t \beta_i}\right)\left(W_E+1\right)+\sigma_{ant}^2 \Bigg).
\end{align}
$Proof:$\\
Based on (\ref{Quser}) and (\ref{Q eaves}), the total received RF power by the users and the eavesdropper can be expanded respectively as 
\begin{align} \label{expandQuser}
P^{EH}_{k}&=(1-\rho_{k})\Bigg(\mathbb{E}\left\lbrace\sum_{i=1}^{K}\lvert\bm{g}_k^H\bm{w}_i\left(s_i+w_{E_i}\right)\rvert^2\right\rbrace+\mathbb{E}\left\lbrace \lvert n_{ant,k}\rvert^2\right\rbrace \Bigg), 
\end{align}
\begin{align} \label{expandQeaves}
P^{EH}_{\text{Eve}}&=(1-\rho_{\text{Eve}})\Bigg(\mathbb{E}\left\lbrace \sum_{i=1}^{K}\lvert\bm{g}_w^H\bm{w}_i\left(s_i+w_{E_i}\right)\rvert^2\right\rbrace +\mathbb{E}\left\lbrace \lvert n_{ant,Eve}\rvert^2\right\rbrace \Bigg).   \quad \blacksquare
\end{align}
\subsection{Achievable Secrecy Rate} 
After power splitting, the $k$'th user and eavesdropper signal for information decoding can be respectively rewritten as
\begin{align} \label{user signal rewrite}
y_{k_{\text{ID}}}&=\sqrt{\rho_k}(\bm{g}_k^H\bm{w}_ks_k+\mathbb{E}\left\lbrace \bm{g}_k^H\bm{w}_k\right\rbrace s_k-\mathbb{E}\left\lbrace \bm{g}_k^H\bm{w}_k\right\rbrace s_k+\dfrac{\sqrt{\eta p_t}\bm{g}_k^H\bm{g}_k w_{E_k}}{\sqrt{M\left(\eta p_t \beta_k+  q_t\beta_w+\sigma^2\right)}}\nonumber \\&+\dfrac{\bm{g}_k^H\left(\bm{N}\bm{\phi}_k^*+\sqrt{\eta q_t}\bm{g}_w \bm{\phi}_w^T\bm{\phi}_k^*\right)}{\sqrt{M\left(\eta p_t \beta_k+ q_t\beta_w+\sigma^2\right)}}w_{E_k}+\bm{g}_k^H \sum_{\begin{smallmatrix}i=1\\i\neq k \end{smallmatrix}}^{K}\bm{w}_i(s_i+w_{E_i})+n_{ant,k})+n_s,
\end{align}
\begin{align}\label{eaves signal rewrite}
y_{\text{Eve}_{\text{ID}}}&=\sqrt{\rho_{Eve}}(\bm{g}_w^H\bm{w}_ks_k+\mathbb{E}\left\lbrace \bm{g}_k^H\bm{w}_k\right\rbrace s_k-\mathbb{E}\left\lbrace \bm{g}_w^H\bm{w}_k\right\rbrace s_k\nonumber\\
&+\bm{g}_w^H \sum_{\begin{smallmatrix}i=1\\i\neq k \end{smallmatrix}}^{K}\bm{w}_i s_i+\bm{g}_w^H \sum_{i=1}^{K}\bm{w}_iw_{E_i}+n_{ant,Eve})+n_s,
\end{align}
where $n_s$ is additional processing noise modeled as $n_s\in \mathcal{C}\mathcal{N}(0,\sigma_s^2)$.

\noindent The achievable secrecy rate is defined as \cite{secrecyformula}
\begin{align} \label{exact secrecy rate}
R_{\text{Secrecy,k}}=\mathbb{E}\left\lbrace [R_{\text{k}}-R_{\text{Eve}}]^{+}\right\rbrace, 
\end{align}
where $R_{\text{k}}$ and $R_{\text{Eve}}$ are the user and eavesdropper achievable rate, respectively.\\
\begin{align} \label{secrecylowerbound}
&\text{R}_{\text{S,bound}}=\dfrac{T-\tau}{T}\log_2\left(1+\dfrac{\rho_kM\eta p_t \beta_k^2/\left(\sigma^2+q_t \beta_w+\eta p_t \beta_k\right)}{\rho_k\left(K\beta_k+(k-1)\beta_kW_E+\dfrac{\beta_k(q_t \beta_w+\sigma^2)}{\eta p_t\beta_k+q_t \beta_w+\sigma^2}+\sigma_{ant}^2 \right)+\sigma_s^2}\right)\nonumber\\
&-\dfrac{T-\tau}{T}\log_2\left(1+\dfrac{\rho_{\text{Eve}}M\eta q_t \beta_w^2/\sigma^2+q_t \beta_w+\eta p_t   \beta_k}{\rho_{\text{Eve}} \left(\left(K\beta_w+\sum_{i=1}^K \dfrac{Mq_t \beta_w^2}{\sigma^2+q_t \beta_w+\eta p_t \beta_i }\right)\left(W_E+1\right)-\dfrac{M q_t \beta_w^2}{\sigma^2+q_t \beta_w+\eta p_t \beta_k}+\sigma_{ant}^2\right)+\sigma_s^2}\right)
\end{align}
\textbf{Theorem 2:} By exploiting MRT precoding and MMSE channel estimation, the $k$th user achievable secrecy rate lower bound can be represented as (\ref{secrecylowerbound}). 

$Proof$: The proof is provided in Appendix A.
\section{asymptotic analysis}
To obtain an analytical insight, a massive MIMO system in which the number of antennas grows sufficiently large is considered. \newline
The asymptotic total RF power received  at the $k$'th user and the eavesdropper can be expressed as
\begin{align}
&P_{k}^{\text{asym}}=(1-\rho_k) \dfrac{M\eta p_t \beta_k^2}{\left(\eta p_t \beta_{k}+q_t \beta_{w}+\sigma^2\right)}\left(W_E+1\right), \label{asymquesr}\\
&P_{Eve}^{\text{asym}}=(1-\rho_{Eve})\sum_{i=1}^{K} \dfrac{M q_t \beta_k^2}{\left(\eta p_t \beta_{k}+q_t \beta_{w}+\sigma^2\right)}\left(W_E+1\right).\label{asymeaves}
\end{align}
As it can be seen, the total RF power received at the user and eavesdropper increase by the number of antennas. Also, the non-linear harvested energies in the asymptotic case are
\begin{align}
Q^{\text{asym}}_{\text{ NL, k}}&=\dfrac{P_{s_k}}{\text{exp}(a\times b)}\times\left[ \dfrac{1+\text{exp}(a\times b)}{1+\text{exp}\left(-a([P^{asym}_{k}-P_{SEN}]^{+}-b)\right)}-1\right](T-\tau),
\end{align}
\begin{align}
Q^{\text{asym}}_{\text{NL, Eve}}&=\dfrac{P_{s_{Eve}}}{\text{exp}(a\times b)}\times\left[ \dfrac{1+\text{exp}(a\times b)}{1+\text{exp}\left(-a([P^{asym}_{Eve}-P_{SEN}]^{+}-b)\right)}-1\right](T-\tau),
\end{align}
It is also seen that the non-linear harvested energy increases by the number of antennas since it is a monotonically increasing function of the total received RF power\cite{energynonelinear}. Due to hardware limitations, the non-linear harvested energy saturates to $P_{s_k}$ for extreme large number of antennas. \\
%Also, the asymptotic secrecy rate lower bound can be expressed as
When the number of antennas grows sufficiently large, the asymptotic secrecy rate increases with the number of antennas and the effect of the eavesdropper can be neglected as shown in (\ref{secrecy rate asymptotic}).
%\newcounter{Mytempeq}
%\begin{figure*}[!t]
%	\normalsize
%	\setcounter{Mytempeq}{\value{equation}}
%	\setcounter{equation}{25}
%	\begin{align} \label{secrecy rate asymptotic}
%	R_{\text{S,bound}}^{\text{asym}}&=\dfrac{T-\tau}{T}\log_2\left(1+M\dfrac{\rho_k\eta p_t \beta_k^2/(\sigma^2+q_t \beta_w+\eta p_t \beta_k)}{\rho_k\big(K\beta_k+(k-1)\beta_kW_E+\dfrac{\beta_k(q_t \beta_w+\sigma^2)}{\eta p_t\beta_k+q_t \beta_w+\sigma^2}+\sigma_{ant}^2 \big)+\sigma_s^2}\right) \nonumber \\
%	&-\dfrac{T-\tau}{T}\log_2\left(1+\dfrac{q_t\beta_w^2}{\left(\sum_{i=1}^{K}\dfrac{q_t\beta_{w}^2}{\eta p_t \beta_{i}+q_t \beta_{w}+\sigma^2}(W_E+1)-\dfrac{ q_t \beta_w^2}{\sigma^2+q_t \beta_w+\eta p_t \beta_k}\right)\left(\eta p_t \beta_{k}+q_t \beta_{w}+\sigma^2\right)}\right)
%	\end{align}
%	\setcounter{equation}{\value{Mytempeq}}
%	\hrulefill
%	\vspace*{4pt}
%\end{figure*} 

\begin{align} \label{secrecy rate asymptotic}
	&R_{\text{S,bound}}^{\text{asym}}=\dfrac{T-\tau}{T}\log_2\left(1+M\dfrac{\rho_k\eta p_t \beta_k^2/(\sigma^2+q_t \beta_w+\eta p_t \beta_k)}{\rho_k\big(K\beta_k+(k-1)\beta_kW_E+\dfrac{\beta_k(q_t \beta_w+\sigma^2)}{\eta p_t\beta_k+q_t \beta_w+\sigma^2}+\sigma_{ant}^2 \big)+\sigma_s^2}\right) \nonumber \\
	&-\dfrac{T-\tau}{T}\log_2\left(1+\dfrac{q_t\beta_w^2}{\left(\sum_{i=1}^{K}\dfrac{q_t\beta_{w}^2}{\eta p_t \beta_{i}+q_t \beta_{w}+\sigma^2}(W_E+1)-\dfrac{ q_t \beta_w^2}{\sigma^2+q_t \beta_w+\eta p_t \beta_k}\right)\left(\eta p_t \beta_{k}+q_t \beta_{w}+\sigma^2\right)}\right)
\end{align}

\section{Achievable Secrecy Rate Maximization}
   Since the achievable secrecy rate is a criterion to assess communication security, the achievable secrecy rate is maximized in this section. The maximization is done subject to the required user harvested energy that guarantees user's proper operation ($Q_{\text{min}}$) and a constraint on eavesdropper's harvested energy for limiting its operation ($Q_{\text{max}}$). The optimization parameters are power splitting ratio ($\rho_k$) and harvested energy allocation factor ($\theta$). Considering the above issues, the resulting optimization problem becomes\\
   \setcounter{equation}{26}
   \begin{subequations}
   	\begin{align}
   	&\underset{ \theta,\rho_k}{\text{maximize}}\quad {R_{\text{Secrecy,k}}} \label{secrecylowerboundobject}\\
   	&\;\mathrm{s.t.} \nonumber\\
   	& \quad \qquad Q^{NL}_{\text{k}}\geq Q_{\text{min}} \label{quserconstraint}\\
   	& \quad \qquad Q^{NL}_{\text{Eve}}\leq Q_{\text{max}} \label{qeavesconstraint}\\
   	&\qquad \; \quad 0\leq\rho_k\leq 1 \label{rhoconstraint}\\
   	&\qquad \; \quad 0\leq\theta\leq1 \label{thetaconstraint}.
   	\end{align}
   \end{subequations}
   Since computation of exact achievable secrecy rate in (\ref{secrecylowerboundobject}) is complex, its lower bound in the previous section can replace the objective function. Also (\ref{qeavesconstraint}) can be replaced with $\theta> \theta_{\text{min}}$, since $\theta$ is in the denominator in (\ref{Q eaves2}) and the eavesdropper's harvested energy is strictly decreasing with $\theta$. Due to the $(1-\rho_k)$ coefficient in (\ref{calculated user energy}), user harvested energy is strictly decreasing with $\rho_k$. According to that and since $\theta> \theta_{\text{min}}$, (\ref{quserconstraint}) can be replaced with $\rho_k <\rho_{k_{max}}$. Hence, the optimization problem in (23) can be reformulated as 
   \begin{subequations}
   	\begin{align}
   	&\underset{ \theta,\rho_k}{\text{maximize}} \quad {R_{\text{S,bound}}}\\
   	&\;\mathrm{s.t.} \nonumber\\
   	&\qquad \qquad 0\leq\rho_k\leq\rho_{k_{\text{max}}} \\
   	&\qquad \qquad \theta_{\text{min}}\leq\theta\leq1. 
   	\end{align}
   \end{subequations}
   In the feasible set of this problem, the objective function is strictly increasing with respect to the two parameters $\theta$ and $\rho_k$  (see appendix C). Hence, we conclude that the optimal value is a point on the border of the feasible set \cite{boyd2004convex}. 
   In other words, to maximize the lower bound of the secrecy rate, it is enough to set both the parameters of  $\theta$ and $\rho_k$ to their maximum, i.e., $\theta=1$ and $\rho_k=\rho_{k_{\text{max}}}$.
   \\
   The harvested energy by the users is used for uplink pilot training and data processing. Since the power consumption of data processing is relatively small, it can be neglected and we can assume $\theta=1$ \cite{massiveSWIPT2018}, \cite{massivemimo2015}, \cite{negligiblepowerconsumption}. Accordingly, the optimal secrecy rate lower bound can be represented as (\ref{optimalsecrecylowerbound}).
   
   \begin{align} \label{optimalsecrecylowerbound}
      	&\text{R}_{\text{S,bound optimal}}=\dfrac{T-\tau}{T}\log_2\left(1+\dfrac{\rho_{k_{\text{max}}}M Q^{NL}_k \beta_k^2/\left(\sigma^2+q_t \beta_w+ Q^{NL}_k \beta_k\right)}{\rho_{k_{\text{max}}}\left(K\beta_k+(k-1)\beta_kW_E+\dfrac{\beta_k(q_t \beta_w+\sigma^2)}{Q^{NL}_k\beta_k+q_t \beta_w+\sigma^2}+\sigma_{ant}^2 \right)+\sigma_s^2}\right)\nonumber\\
      	&-\dfrac{T-\tau}{T}\log_2\left(1+\dfrac{\rho_{\text{Eve}}M\eta q_t \beta_w^2/\sigma^2+q_t \beta_w+ Q^{NL}_k \beta_k}{\rho_{\text{Eve}} \left(\left(K\beta_w+\sum_{i=1}^K \dfrac{Mq_t \beta_w^2}{\sigma^2+q_t \beta_w+ Q^{NL}_k \beta_i }\right)\left(W_E+1\right)-\dfrac{M q_t \beta_w^2}{\sigma^2+q_t \beta_w+ Q^{NL}_k \beta_k}+\sigma_{ant}^2\right)+\sigma_s^2}\right)
  \end{align}

%   \newcounter{Mytempeqn}
%   \begin{figure*}[!t]
%   	\normalsize
%   	\setcounter{Mytempeq}{\value{equation}}
%   	\setcounter{equation}{28}
%   	\begin{align} \label{optimalsecrecylowerbound}
%   	\text{R}_{\text{S,bound optimal}}&=\dfrac{T-\tau}{T}\log_2\left(1+\dfrac{\rho_{k_{\text{max}}}M Q^{NL}_k \beta_k^2/\left(\sigma^2+q_t \beta_w+ Q^{NL}_k \beta_k\right)}{\rho_{k_{\text{max}}}\left(K\beta_k+(k-1)\beta_kW_E+\dfrac{\beta_k(q_t \beta_w+\sigma^2)}{Q^{NL}_k\beta_k+q_t \beta_w+\sigma^2}+\sigma_{ant}^2 \right)+\sigma_s^2}\right)\nonumber\\
%   	&-\dfrac{T-\tau}{T}\log_2\left(1+\dfrac{\rho_{\text{Eve}}M\eta q_t \beta_w^2/\sigma^2+q_t \beta_w+ Q^{NL}_k \beta_k}{\rho_{\text{Eve}} \left(\left(K\beta_w+\sum_{i=1}^K \dfrac{Mq_t \beta_w^2}{\sigma^2+q_t \beta_w+ Q^{NL}_k \beta_i }\right)\left(W_E+1\right)-\dfrac{M q_t \beta_w^2}{\sigma^2+q_t \beta_w+ Q^{NL}_k \beta_k}+\sigma_{ant}^2\right)+\sigma_s^2}\right)
%   	\end{align}
%   	\setcounter{equation}{\value{Mytempeq}}
%   	\hrulefill
%   	\vspace*{4pt}
%   \end{figure*} 
\section{Numerical Results and Analysis}
A single cell SWIPT enabled massive MIMO system with four users and one active eavesdropper is simulated. The large scale fading are modeled as $\beta_{k}=10^{-3}d_k^{-3}$ and $\beta_{w}=10^{-3}d_w^{-3}$ for the users and eavesdropper, respectively \cite{massiveSWIPT2017}, where $d_k \sim \mathcal{U}[10,20]$ (meters) and  $d_w \sim \mathcal{U}[10,20] $ (meters) denote the $k$th user's and eavesdropper's distance from the BS \cite{massiveSWIPT2018}. The simulation parameters are shown in Table \ref{tableparameters}.\newline
\begin{table} [!h]
	\centering 
	\caption{simulation parameters}
	\label{tableparameters}
	\begin{tabular}{|c|c|}
		\hline
		Parameter & Value \\
		\hline
		Coherence time: T & 5 ms  \\
		Pilot sequence length: $\eta$ & 4	\\
		Eavesdropper power splitting ratio: $\rho_{\text{Eve}}$ & 0.5\\
		BS noise power in the training phase: $\sigma^2$ & -90 dBm \\ 
		Receiver AWGN noise power: $\sigma_{\text{ant}}^2$& -70 dBm  \\
		Processing noise power: $\sigma_s^2$ &-50 dBm \\ 
		EH circuit specification: $\text{a}$ & 150\\
		EH circuit specification: $\text{b}$ & 0.014\\
		Maximum harvested power at the user: $P_{s_k}$& -40dBm\\
		Maximum harvested power at the eavesdrpper: $P_{s_{Eve}}$& -40dBm\\
		EH sensitivity: $P_{SEN}$ & 0.024 mW\\
		Eavesdropper harvested energy allocation factor: $\zeta$ &0.5 \\
		\hline
	\end{tabular}
\end{table}
\begin{figure}[!h]
	%\hspace{-0.5cm}
	\centering
	\includegraphics[scale=0.435,,trim=3.2cm 0.6cm 15.2cm 0.6cm ,clip]{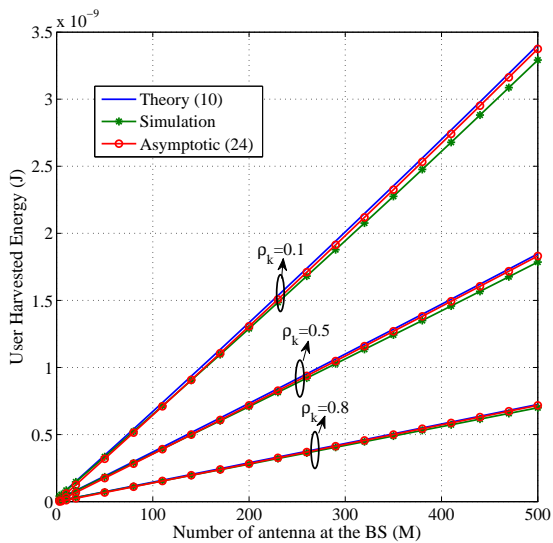}
	\caption{Average user's harvested energy versus the number of antennas at the BS, where $\rho_k=[0.1, 0.5 , 0.8]$ and $\theta=0.7$.}
	\label{fig:userharvestedenergy}
\end{figure}
Fig. \ref{fig:userharvestedenergy} shows the average user's harvested energy versus the number of antennas at the BS for $\rho_k=[0.1, 0.5 , 0.8]$. It can be seen that simulated and theoretical results for the user's harvested energy are very close to each other. Also, the results indicate the accuracy of the asymptotic expression for the user's harvested energy. It is observed that the amount of harvested energy increases linearly with the number of antennas and decreases with the user power splitting ratio.

\begin{figure}
	%\hspace{-0.4cm}
	\centering
	\includegraphics[scale=0.44,trim=3.2cm 0.7cm 15.7cm 0.6cm ,clip]{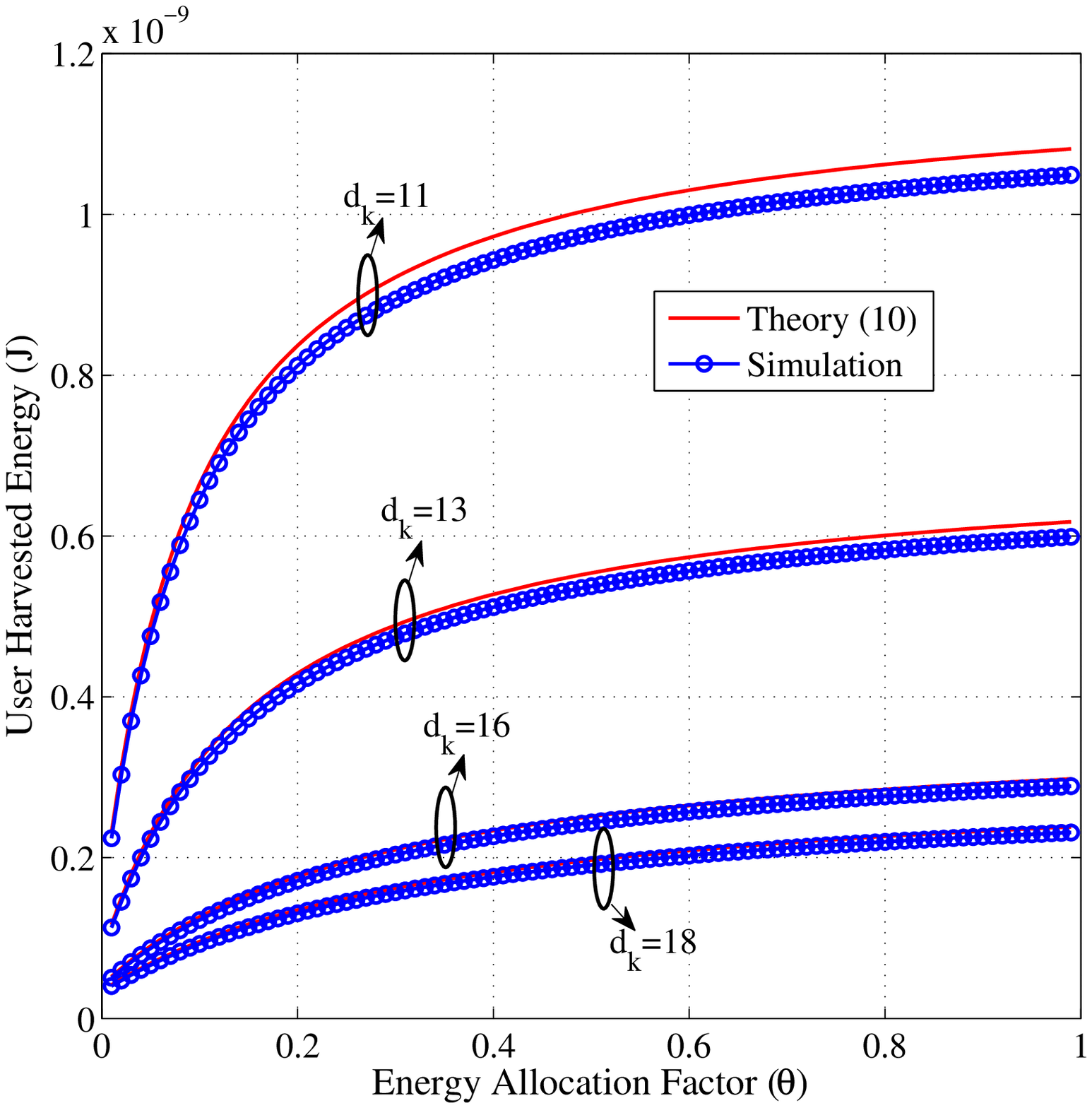}
	\caption{Average user's harvested energy versus the fraction of energy allocated to pilot training by the user. The users and eavesdropper distance from the BS are $d_k=[11, 13, 16 ,18]$, $d_w=15$, respectively and $\rho_k=0.4$.}
	\label{fig:quser}
\end{figure}
\begin{figure}[!h]
	%\hspace{-0.42cm}
		\centering
	\includegraphics[scale=0.42,trim=3.2cm 0.7cm 14.5cm 1cm,clip]{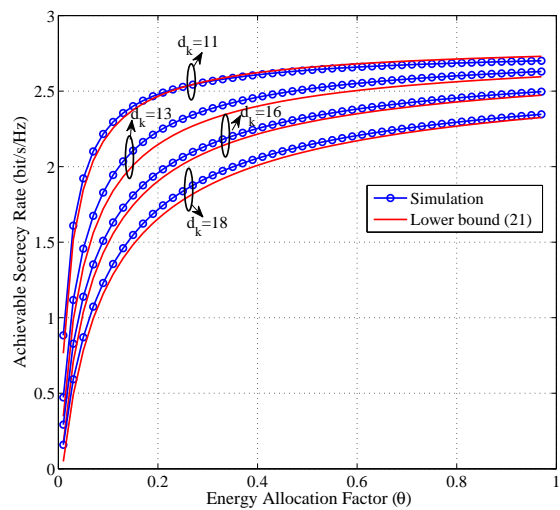}
	\caption{User's achievable secrecy rate versus the fraction of energy allocated to pilot training by the user. The users and eavesdropper distance from the BS are $d_k=[11, 13, 16 ,18]$, $d_w=15$, respectively and $\rho_k=0.4$.}
	\label{fig:ratethetausers}
\end{figure}
\begin{figure}[!h]
	%	\hspace{-0.4cm}
		\centering
	\includegraphics[scale=0.39,trim=3.9cm 0.6cm 13cm 0.3cm ,clip]{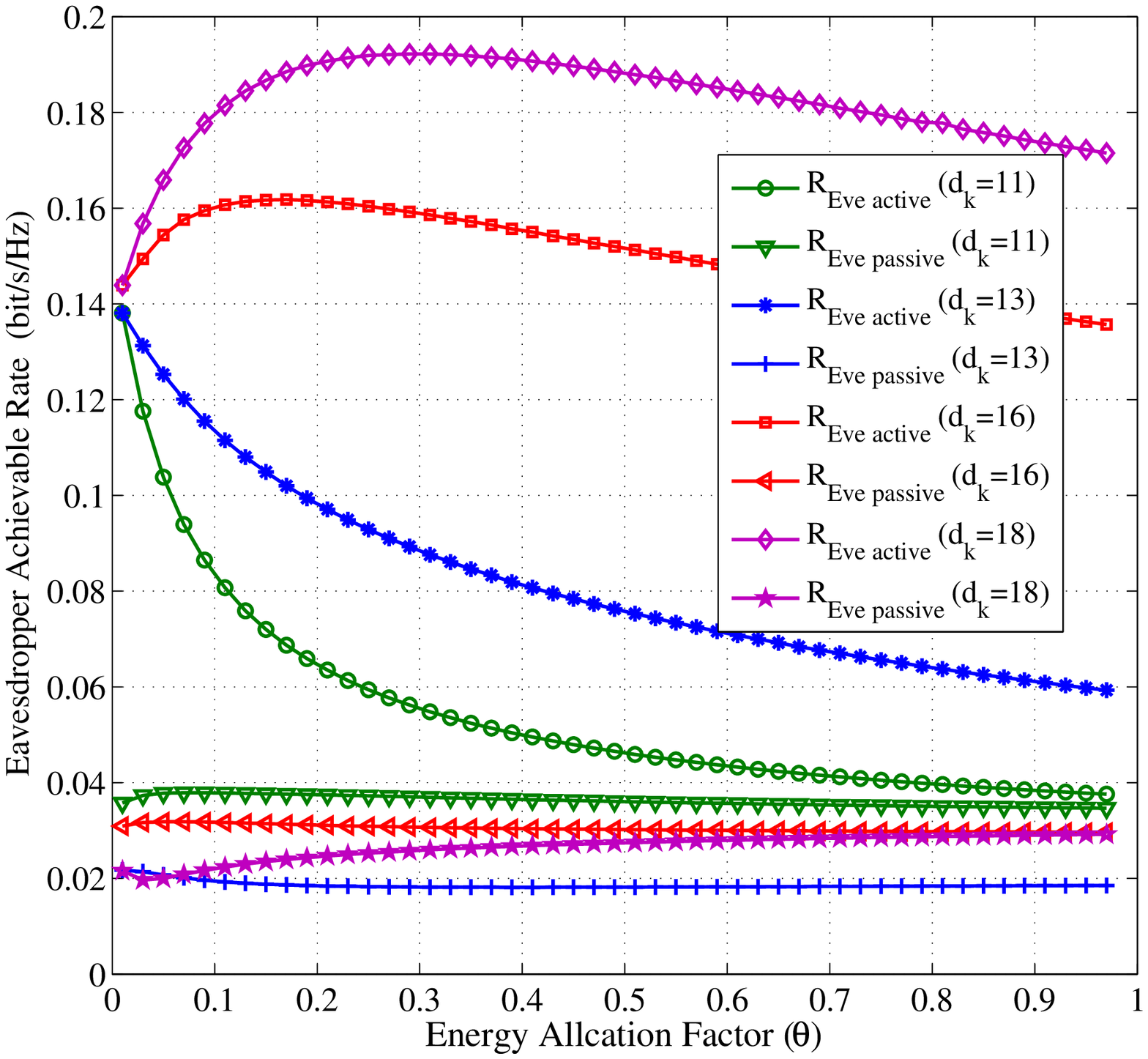}
	\caption{Average achievable rate at the eavesdropper versus the fraction of energy allocated to pilot training by the user. The users, active eavesdropper and passive eavesdropper distance from the BS are $d_k=[11, 13, 16 ,18]$, $d_w=15$ and $d_{\text{passive}}=15$, respectively and $\rho_k=0.4$.}
	\label{fig:active-passive2}
\end{figure}
\begin{figure}[!h]
	%	\hspace{-0.4cm}
	\centering
	\includegraphics[scale=0.43,trim=3.5cm 0.67cm 15.6cm 1.1cm ,clip]{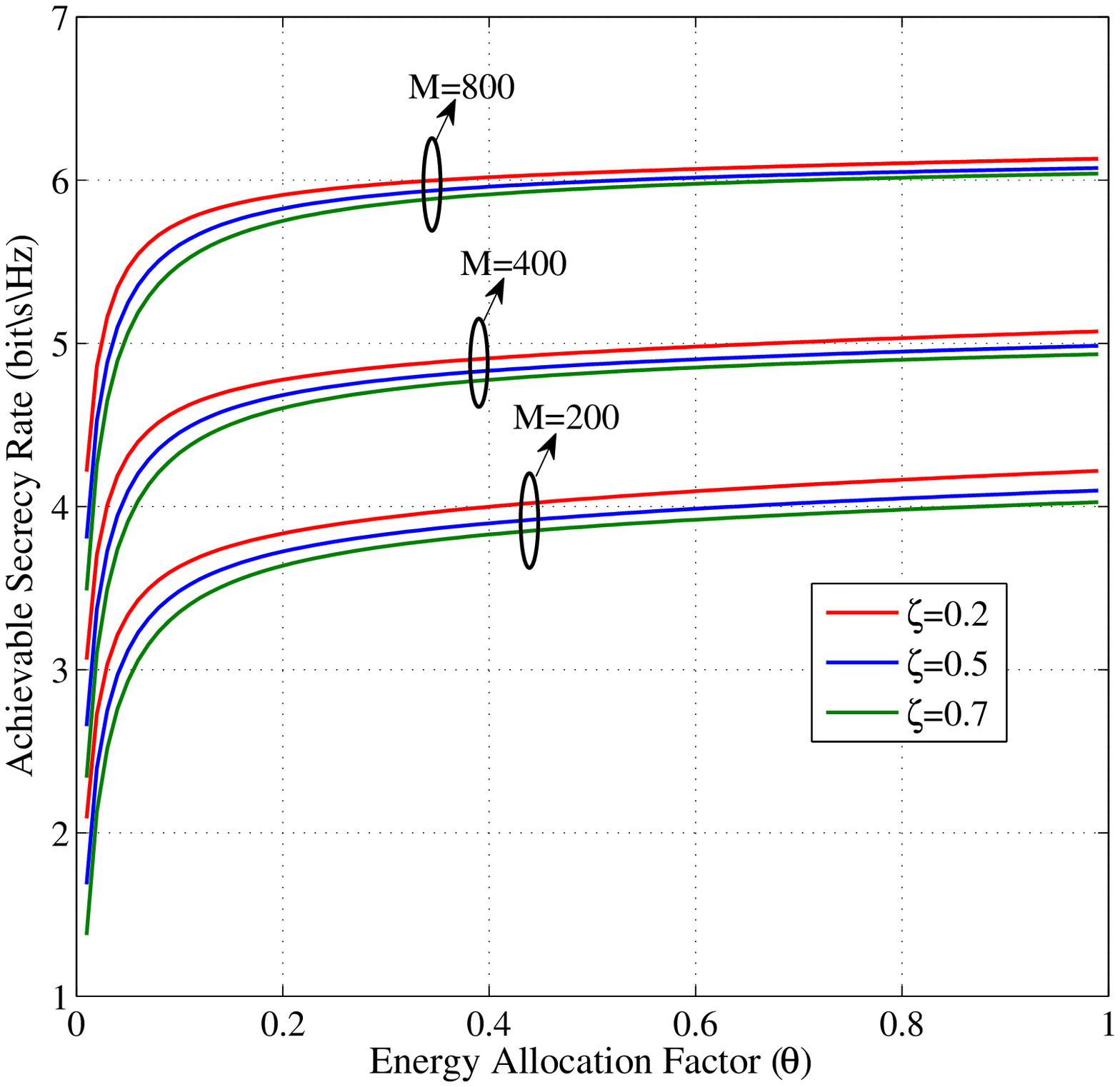}
	\caption{Average achievable secrecy rate versus the fraction of energy allocated to pilot training by the user. The user and active eavesdropper distances from the BS are $d_k=13$ and $d_w=15$, respectively. User power splitting ratio and eavesdropper power splitting ratio are set  $\rho_k=0.4$ and $\zeta=[0.2 \quad 0.5 \quad 0.7]$ for the number of antennas at BS $M=[200 \quad 400 \quad 800]$.}
	\label{fig:eavespowersplitting}
\end{figure}
\begin{figure}[!h]
	%\hspace{0.4cm}
	\centering
	\includegraphics[scale=0.41,trim=3.1cm 0.7cm 15cm 1cm ,clip]{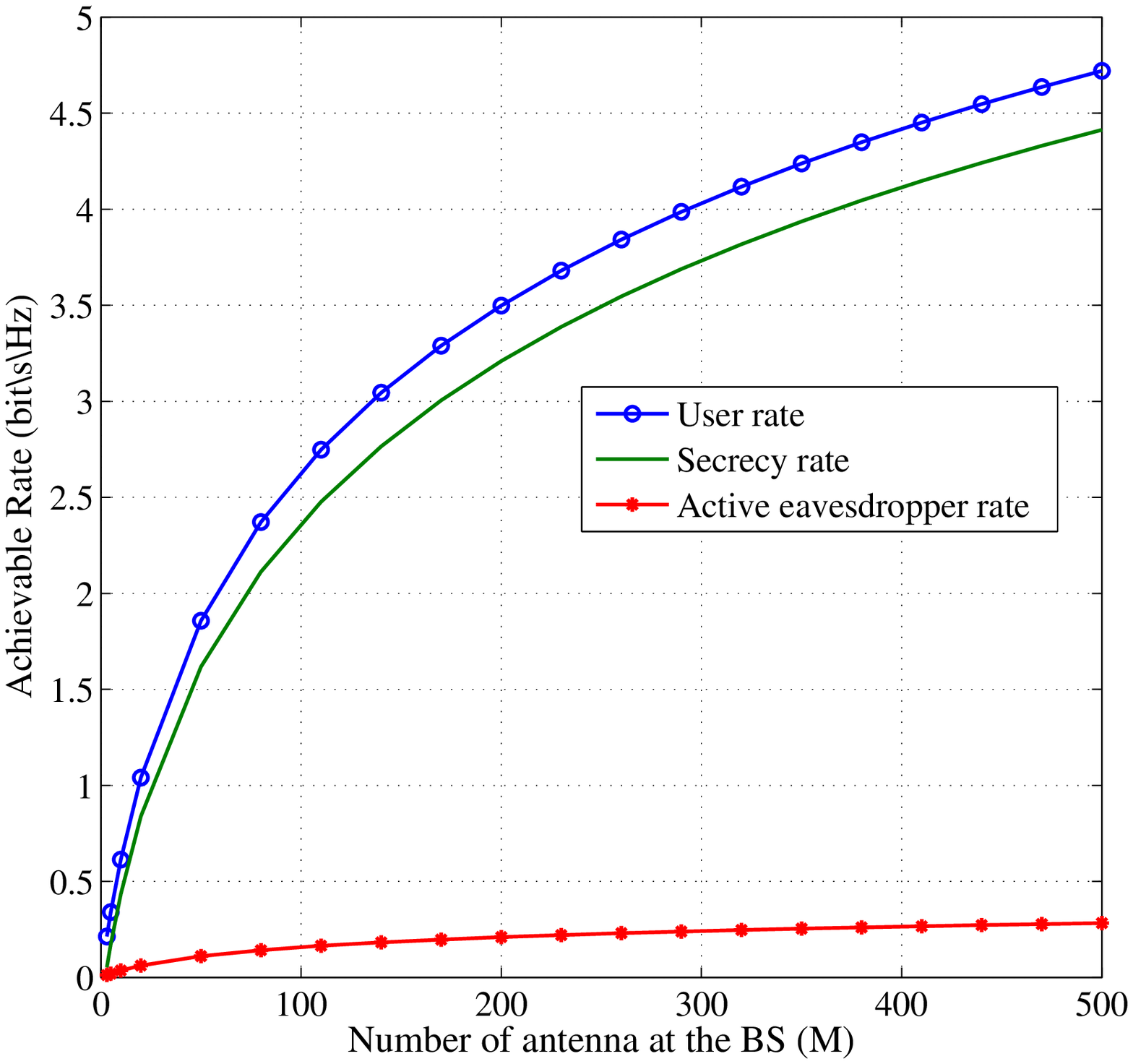}
	\caption{Achievable Rates versus the number of antennas at the BS. The user and eavesdropper distances from the BS are $d_k=13$ and $d_w=15$. $\rho_k=0.4$ and $\theta=0.7$ }
	\label{fig:Mrate}
\end{figure}
%\begin{figure}[!h]
%%	\hspace{-.5cm}
%\centering
%\includegraphics[scale=0.45,trim=4.35cm 0.7cm 5.3cm 0.9cm ,clip]{ENsignal.eps}
%	\caption{Achievable secrecy rate with exploiting energy signal versus number of anttenas}
%	\label{fig:energysignal}
%\end{figure}

Fig. \ref{fig:quser} illustrates the average user's harvested energy versus the fraction of energy allocated to pilot training by the user. This figure shows that the amount of harvested energy increases by the fraction of energy allocated to the pilot training. In fact, increasing the allocated energy to pilot training, channel estimation accuracy is improved and the BS provides more concentrated beams towards the users. Thus energy is transferred more efficiently and the amount of harvested energy by the user is increased. The figure also shows a good agreement between the theoretical and simulated results. 

Fig. \ref{fig:ratethetausers} shows the user's achievable secrecy rate versus the fraction of energy allocated to pilot training by the user. It can be observed that the achievable secrecy rate increases with the allocated energy to pilot training, due to more accurately beamformed information signals to the users. Furthermore, the simulated achievable secrecy rate and the obtained lower bound are close to each other.

Fig. \ref{fig:active-passive2} shows the active and passive eavesdropper's achievable rates versus the fraction of energy allocated to the users' uplink pilot training. Here by passive eavesdropper, we mean an eavesdropper who does not transmit any signals and only harvests energy and decodes the information at the downlink. This figure shows a comparison between the two cases of active and passive eavesdropper and the effect of massive MIMO in these two cases. As can be observed, for users that are located closer to the BS than the active eavesdropper, channel estimation and beamforming are more accurate. Thus, the eavesdropper's achievable rate decreases by increasing the energy allocated to pilot training. For users where the active eavesdropper is located closer to the BS, the eavesdropper's achievable rate first increases due to stronger pilot training by the eavesdropper and then decreases due to increasing the power of user pilot training. Also, it can be seen that the passive eavesdropper's achievable rate is extremely low, and increasing $\theta$ does not affect it. This is due to that the passive eavesdropper does not attend in the pilot training.

Fig. \ref{fig:eavespowersplitting} indicates that the eavesdropper can improve its rate and reduce the secrecy rate by allocating more power to its pilot phase. Also, it can be seen that by increasing the number of antennas at the BS (i.e. enabling narrower beams toward the users), allocating more power to the pilot phase by the eavesdropper will not improve its rate and the secrecy rate does not change any more.

Fig. \ref{fig:Mrate} shows the achievable rates versus the number of antennas at the BS. It can be seen that in small number of antennas ( i.e. when $M$ is small which is related to the conventional MIMO), the user's and eavesdropper's achievable rate increase by the number of antennas. However, when the number of antennas grows very large  ( i.e. when $M$ is large which is related to the Massive MIMO), the eavesdropper rate is limited to a constant value while the user's rate still increases with the number of antennas. In other words, this figure shows a comparison between the results of the proposed scheme in two cases of conventional MIMO and massive MIMO systems. 

%Fig. \ref{fig:energysignal} shows the effect of exploiting energy signal on security of transmission.
\section{conclusions}
In this paper, a secure SWIPT system exploiting power splitting in the downlink of a multi user massive MIMO system with uplink pilot training was investigated. To assess the security of communication a lower bound for the achievable secrecy rate was derived. Based on the derived lower bound, the power splitting ratio and the fraction of harvested energy allocated to uplink pilot training were obtained in order to maximize the achievable secrecy rate subject to the minimum harvested energy required for the user and the maximum harvested energy for the eavesdropper to restrict its performance. The numerical results verify the accuracy of the obtained secrecy rate lower bound. It is revealed that massive MIMO noticeably can improve communication security by transmitting data towards users via narrow beams.\\ In this paper, it is assumed that the active eavesdropper channel and the users' channels are independent. Assuming various degree of correlations between the channels of eavesdropper and the users, it is interesting to investigate how this correlation could affect the eavesdropping and 
if massive MIMO can prevent information leakage also in this case. It seems likely that correlation between the channels of the eavesdropper and the users would help the eavesdropper to more easily eavesdrop information and harvest more energy, since redirecting the signal beam toward itself might be easier in this case. 
%Also, transmitting energy signal can improve communication security by confounding the eavesdropper. 
\section*{Appendix A}
\section*{proof of theorem 1}
A lower bound on achievable secrecy rate (\ref{exact secrecy rate}) is obtained as below.
\setcounter{equation}{29} 
\begin{align} \label{secrecy rate bound1}
\mathbb{E}\{[R_{\text{k}}-R_{\text{Eve}}]^{+}\}& \stackrel{a}{=}\mathbb{E}\{ \max(\left( R_{\text{k}}-R_{\text{eaves}},0\right) \}\nonumber\\
& \stackrel{b}{\geq}\max\left(\mathbb{E}\left\lbrace R_{\text{k}}-R_{\text{Eve}}\right\rbrace ,0\right)\nonumber\\
& \stackrel{c}{\geq}\mathbb{E}\left\{ R_{\text{k}}-R_{\text{Eve}}\right\},
\end{align}
where in (a) the achievable secrecy rate is written in another form, (b) is because of Jensen's inequality and (c) is because the maximum of two values is greater than or equal to  each of them.\newline 
According to (\ref{user signal rewrite}) and (\ref{eaves signal rewrite}), user and eavesdropper rate can be expressed as
\begin{align} \label{lower bound1 user}
R_{\text{k}}=\mathbb{E}\left\lbrace \dfrac{T-\tau}{T}\log_2\left(1+\nonumber
\dfrac{\rho_k\left|\mathbb{E}\{\bm{g}_k^H\bm{w}_k\}s_k\right|^2}{U}\right)\right\rbrace, 
\end{align}
where $U$ is
\begin{align}
U=&\rho_k\bigg(\sum_{\begin{smallmatrix}i=1\\i\neq k \end{smallmatrix}}^{K}\left| \bm{g}_k^H\bm{w}_i\right| ^2(W_E+1)+\left| \bm{g}_k^H\bm{w}_k-\mathbb{E}\{\bm{g}_k^H\bm{w}_k\}\right|^2 \nonumber\\
&\left|\dfrac{\bm{g}_k^H(\bm{N}\bm{\phi}_k^*+\sqrt{\eta q_t}\bm{g}_w \bm{\phi}_w^T\bm{\phi}_k^*)}{\sqrt{M(\eta p_t \beta_k+ q_t\beta_w+\sigma^2)}}\right| ^2W_{E}+\sigma_{ant}^2\bigg)+\sigma_s^2.
\end{align} 
Although the user knows the energy signal which is transmitted by the BS, it can not remove it completely from the received signal due to imperfect channel state information (CSI). The third term in $U$ refers to this issue.    
\begin{align} \label{lower bound1 eaves}
&R_{\text{Eve}}=\mathbb{E}\bigg\{\dfrac{T-\tau}{T}\log_2\bigg(1+\dfrac{\rho_{E}|\mathbb{E}\left\lbrace\bm{g}_{w}^H\bm{w}_{k}\right\rbrace|^{2} }{Z}\bigg)\bigg\},
\end{align}
where $Z$ is \newline
\begin{align}
Z=&\rho_{\text{Eve}}\left(\sum_{i=1}^{N}\left|\bm{g}_w \bm{w}_i\right|^2(W_E+1)-\left|\mathbb{E}\left\lbrace\bm{g}_{w}^H\bm{w}_{k}\right\rbrace\right|^{2}+\sigma_{ant}^2\right)+\sigma_s^2.
\end{align} 
According to Jensen inequality a lower bound for (\ref{lower bound1 user}) is obtained as 
\begin{align} \label{user bound1}
&\mathbb{E}\left\lbrace \dfrac{T-\tau}{T}\log_2\left(1+
\dfrac{\rho_k\left|\mathbb{E}\left\lbrace \bm{g}_k^H\bm{w}_k\right\rbrace s_k\right|^2}{U}\right)\right\rbrace  
\geq \dfrac{T-\tau}{T}\log_2\bigg(1+\dfrac{\rho_k\left|\mathbb{E}\{\bm{g}_k^H\bm{w}_k\}\right|^2}{\mathbb{E}\{U\}}\bigg).
\end{align}
Also an upper bound for (\ref{lower bound1 eaves}) can be obtained according to the generalized Jensen's inequality \cite{jensengao2017bounds}
\begin{align} \label{eaves inequality}
&\mathbb{E}\left\lbrace \frac{T-\eta}{T}\log_2\left(1+\dfrac{\rho_{\text{Eve}}|\mathbb{E}\left\lbrace\bm{g}_{w}^H\bm{w}_{k}\right\rbrace|^{2} }{Z}\right)\right\rbrace  \leq  frac{T-\eta}{T}\log_2\left(1+\dfrac{\rho_{\text{Eve}}|\mathbb{E}\left\lbrace\bm{g}_{w}^H\bm{w}_{k}\right\rbrace|^{2} }{\mathbb{E}\lbrace Z\rbrace}\right)+C \sigma_ Z^{2},
\end{align}
where $ C=\dfrac{2\mu+1}{4(\mu^{2}+\mu)}$ and $\mu \in [1,\infty]$. By assuming $\mu=1$, the maximum value of $C$ is $C=3/8$. Also $\sigma_Z^2$ is defined as
\begin{align} \label{EZ2-E2Z}
\sigma_Z^2&=\mathbb{E}\left\lbrace Z^2\right\rbrace -\mathbb{E}\left\lbrace Z\right\rbrace ^2=\rho_{\text{Eve}}^2\mathbb{E}\left\lbrace \left(\sum_{i=1}^{N}|\bm{g}_{w}^H \bm{w}_i|^2(W_E+1)^2 \right)^2\right\rbrace -\rho_{\text{Eve}}^2\left(\mathbb{E}\left\lbrace \sum_{i=1}^{N}|\bm{g}_{w}^H \bm{w}_i|^2(W_E+1)\right\rbrace \right)^2
\end{align}
Some useful equations for calculating $\sigma_Z^2$ are provided in appendix B.

By substituting the result of expectation from appendix B, $\sigma_Z^2$ in (\ref{EZ2-E2Z}) is obtained as
\begin{align}
\sigma_Z^2&=\rho_{\text{Eve}}^2\bigg(\big(2M \beta_ w^2 \sum_{i=1}^{K}\frac{\eta^ 2 p_t^2K(K+1) \beta_ i^2+2 \eta p_t \sigma ^2 \beta_ i}{M^2(\eta p_t \beta_ i+q_t \beta_ w+\sigma^ 2)^2} \nonumber \\
&+ (M^2-M) \beta _w^2 \sum_{i=1}^K \frac{ \eta^ 2 p_t^2\beta_{i}^{2}+2 \eta p_t \sigma^ 2 \beta_ i}{M^2(\eta p_t \beta_ i+q_t \beta_ w+\sigma^ 2)^2}\nonumber\\
&+q_t(M+1)(M+2)(6\sigma^2+(M+3)q_t\beta_w)\big)\nonumber\\
&+2K\delta\eta p_tM(M+1)(\sigma^2+(M+2)q_t \beta_w)\beta_w^2\sum_{i=1}^{K} \beta_{i}^2\bigg)\nonumber\\
&(W_E+1)^2+K^2\delta M(\dfrac{M(5M+11)\sigma^2\beta_w^2}{4})-(K\beta_w\nonumber\\
&+\sum_{i=1}^{K}\dfrac{Mq_t \beta_w^2}{\eta p_t \beta_i+q_t \beta_w+\sigma^2})^2(W_E+1)^2.
\end{align}
As $C\sigma_Z^2$ is negligible compared to \newline
$\frac{T-\eta}{T}\log_2\left(1+\dfrac{\rho_{\text{Eve}}|\mathbb{E}\left\lbrace\bm{g}_{w}^H\bm{w}_{k}\right\rbrace|^{2} }{\mathbb{E}\lbrace Z\rbrace}\right)$, (\ref{eaves inequality}) can be written as
\begin{align} \label{eaves bound2}
&\mathbb{E}\left\lbrace \frac{T-\eta}{T}\log_2\left(1+\dfrac{\rho_{\text{Eve}}|\mathbb{E}\left\lbrace\bm{g}_{w}^H\bm{w}_{k}\right\rbrace|^{2} }{Z}\right)\right\rbrace  \leq \frac{T-\eta}{T}\log_2\left(1+\dfrac{\rho_{\text{Eve}}|\mathbb{E}\left\lbrace\bm{g}_{w}^H\bm{w}_{k}\right\rbrace|^{2} }{\mathbb{E}\lbrace Z\rbrace}\right).
\end{align}
By substituting obtained bounds, (\ref{user bound1}) and (\ref{eaves bound2}) into (\ref{secrecy rate bound1}), the achievable secrecy rate lower bound can be expressed as
\begin{align} \label{secrecy rate bound2}
\mathbb{E}\left\lbrace [R_{\text{k}}-R_{\text{Eve}}]^{+}\right\rbrace &\geq \dfrac{T-\tau}{T}\log_2\left(1+\dfrac{\rho_k\left|\mathbb{E}\{\bm{g}_k^H\bm{w}_k\}\right|^2}{\mathbb{E}\{U\}}\right)
-\frac{T-\eta}{T}\log_2\left(1-\dfrac{\rho_{\text{Eve}}|\mathbb{E}\left\lbrace\bm{g}_{w}^H\bm{w}_{k}\right\rbrace|^{2} }{\mathbb{E}\lbrace Z\rbrace}\right).
\end{align}
\begin{figure}
	%\hspace{-0.25 cm}
     \centering
	\includegraphics[scale=.41,trim=3.6cm 0.5cm 4cm 0.8cm,clip]{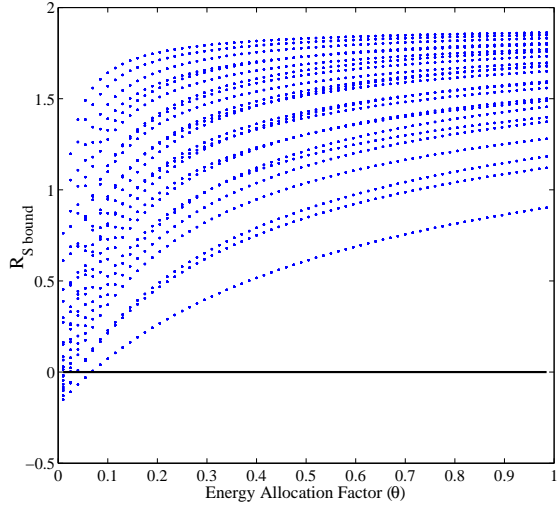}
	\caption{Positivity of secrecy rate lower bound. The secrecy rate lower bound versus harvested energy allocation factor is plotted for various locations of the users and the active eavesdropper. $d_k \sim \mathcal{U}[10,20]$ (meters) and  $d_w \sim \mathcal{U}[10,20] $ (meters). }
	\label{fig:positiveratesecrecy}
\end{figure}
Since the achievable secrecy rate is positive, the positivity of the obtained lower bound is substantial for replacing achievable secrecy rate with its lower bound in (\ref{secrecylowerboundobject}). Lower bound positivity or
negativity is not obvious and also determining its positivity or negativity is not simple. Fig. \ref{fig:positiveratesecrecy}  shows the secrecy rate lower bound for various locations of the users and the eavesdropper. It is almost always positive unless the eavesdropper is located near the BS and the other users are close to the desired user and interference is strong. The secrecy rate lower bound is negative when $\theta<0.05$. The negativity of the lower bound can be ignored due to its scarce occurrence and almost always $\theta>0.05$.
\section*{Appendix B}
In this appendix the first term of (\ref{EZ2-E2Z}) is calculated.
\begin{align}\label{gw wi2}
\mathbb{E}\left\lbrace ( \sum_{i=1}^{N}|\bm{g}_{w}^H \bm{w}_i|^2 )^2\right\rbrace =\mathbb{E}\left\lbrace (\bm{g}_{w}^H(A+B)\bm{g}_{w})^2\right\rbrace 
\end{align}
Here, $A$ and $B$ are defined respectively as
\begin{align}
A=\sum_{i=1}^{K}\dfrac{\eta p_t g_{i} g_{i}^H+g_i(N\phi_ i^*)^H+(N\phi_ i^*) g_i ^H}{M(\eta p_t \beta_ i+q_t \beta_ w+\sigma^ 2)},
\end{align}
\begin{align}
B=&K\big(\eta q_t g_w \phi_ w ^T \phi_ 1^*(\phi_ w ^T \phi_ 1^*)^H g_w^H+2 \sqrt{\eta q_t}N \phi_ 1 ^*(\phi_ w^T\phi_ 1^*)^H g_w^H+N \phi_ 1 ^*(N \phi_ 1 ^*)^H\big)
\sum_{i=1}^{K}\dfrac{1}{M(\eta p_t \beta_ i+q_t \beta_ w+\sigma^ 2)}.
\end{align}
Furthermore,
\begin{align} \label{sumgigi}
&\mathbb{E}\left\lbrace (g_w^H A g_w)(g_w^H A^H g_w)\right\rbrace \nonumber\\
&=\mathbb{E}\left\lbrace  \sum_{ij} A_{ij} g_w(i)^* g_w(j) \sum _{m,n} A_{mn} g_w(m) g_w(n)^* \right\rbrace  \nonumber \\& =\mathbb{E}_{A}\left\lbrace\mathbb{E}_{g_w}\left\lbrace\sum_{i,j}^{M} |A_{i,j}|^2 |g_w(i)|^2 |g_w(j)|^2 \;\vert A\right\rbrace\right\rbrace \nonumber\\
& =\mathbb{E}_{A}\Bigg\lbrace\mathbb{E}_{g_w}\Bigg\lbrace \sum_{\begin{smallmatrix} i=1\\i=j \end{smallmatrix}}^{M} |A_{ii}|^2 |g_w(i)|^4 +\sum_{i\neq j}^{M}|A_{ij}|^2 |g_w(i)|^2 |g_w(j)|^2 \; |A\Bigg\rbrace\Bigg\rbrace,
\end{align}
\begin{align} \label{Aii}
\mathbb{E}\lbrace A_{ii}\rbrace &=\mathbb{E}\lbrace\sum_{k=1}^k (|g_k(i)|^2+2g_k(i)^*N(i))\sum_{l=1}^k (|g_l(i)|^2+2N(i)^*g_l(i))\rbrace\nonumber\\
&=\mathbb{E}\lbrace \sum_{\begin{smallmatrix}k=1\\k=l \end{smallmatrix}}^{K} \eta^ 2 p_t^2|g_k(i)|^4+\sum_{k,l} |g_k(i)|^2 |g_l(i)|^2 +4\sum_{k=1}^K |g_k(i)|^2 N(i)^2 \rbrace \nonumber \\&=\eta^ 2 p_t^2 K(K+1) \beta_ i^2+4  \eta p_t \sigma^ 2 \beta_ i,
\end{align}
\begin{align} \label{Aij}
\mathbb{E}\lbrace A_{ij}\rbrace &=\mathbb{E}\bigg\{\sum_{k=1}^{K}( g_k(i)^*g_k(j)+g_k(i)^*N(j)+N(i)^*g_k(j)) .\sum_{l=1}^K( g_l(j)^*g_l(i)+N(j)^*g_l(i)+g_l(j)^*N(i))\bigg\} \nonumber\\
&=\mathbb{E}\lbrace\sum_{k=1}^{K}(\eta^ 2 p_t^2|g_k(i)|^2 |g_k(j)|^2+\eta p_t|g_k(i)|^2|N(j)|^2+\eta p_t |N(i)|^2|g_k(j)|^2)\rbrace\nonumber \\
&=\sum_{i=1}^K   \eta^ 2 p_t^2 \beta_ i^2+ 2\eta p_t \sigma^ 2 \beta_ i, 
\end{align}
and
\begin{align} \label{gw4}
\mathbb{E}\left\lbrace |g_w(i)|^4\right\rbrace&=\mathbb{E}\left\lbrace (g_{\text{w,re}}^2+g_{\text{w, im}}^2)^2\right\rbrace \nonumber\\
&=\mathbb{E}\lbrace{g_{\text{w,re}}^4+g_{\text{w, im}}^4+2g_{\text{w,re}}^2 g_{\text{w, im}}^2}\rbrace \nonumber \\&=2 \beta_ w^2.
\end{align}
By substituting (\ref{gw4}), (\ref{Aii}) and (\ref{Aij}) into (\ref{sumgigi})
\begin{align} \label{gwAA}
\mathbb{E}\lbrace{(g_w^H A g_w)(g_w^H A^H g_w)}\rbrace &=2M \beta_ w^2 \sum_{i=1}^{K}\frac{\eta^ 2 p_t^2K(K+1) \beta_ i^2+4 \eta p_t \sigma ^2 \beta_ i}{M^2(\eta p_t \beta_ i+q_t \beta_ w+\sigma^ 2)^2} \nonumber \\&+ (M^2-M) \beta _w^2 \sum_{i=1}^K \frac{ \eta^ 2 p_t^2\beta_ i+2 \eta p_t \sigma^ 2 \beta_ i}{M^2(\eta p_t \beta_ i+q_t \beta_ w+\sigma^ 2)^2}
\end{align}
\begin{align}\label{gwBB}
&\mathbb{E}\lbrace(g_w^H B g_w)(g_w^H B^H g_w)\rbrace=K^2\delta(\mathbb{E}\lbrace |g_w^H N\phi_ 1^*(N\phi_ 1^*)^H g_w|^2\rbrace \nonumber\\
&+\mathbb{E}\lbrace|\eta q_t g_w^H g_w\phi_w^T\phi_1^*(\phi_w^T\phi_1^*)^H g_w^H g_w|^2\rbrace \nonumber \\
&+ 4\mathbb{E}\lbrace|\sqrt{\eta q_t}g_w^H N\phi_1^*(\phi_w^T\phi_1^*)^H g_w^H g_w|^2\rbrace \nonumber \\
& +2\mathbb{E}\{\eta q_t g_w^H N\phi_ 1^*(N\phi_ 1^*)^H g_wg_w^H g_w\phi_w^T\phi_1^*(\phi_w^T\phi_1^*)^H g_w^H g_w\} \nonumber\\
& + 4 \mathbb{E}\{g_w^H N\phi_1^*(N\phi_1^*)^H g_w \sqrt{\eta q_t}g_w^H N\phi_1^*(\phi_w^T\phi_1^*)^Hg_w^Hg_w\}  \nonumber \\
& +4\mathbb{E}\{\eta q_t \sqrt{\eta q_t}g_w^H N\phi_1^*(\phi_w^T\phi_1^*)^Hg_w^Hg_w g_w^H g_w\phi_w^T\phi_1^*(\phi_w^T\phi_1^*)^H\nonumber\\
& g_w^H g_w \})=\delta(6q_t\mathbb{E}_{g_w}\{\mathbb{E}_{N}\{g_w^HN\phi_1^*(N\phi_1^*)^H g_w \;\left| g_w^Hg_w\right|^2\; \nonumber\\&|g_w \}\}
+\mathbb{E}\{\left| g_w^HN\phi_1^*\right|^4\}+q_t\mathbb{E}\{\left| g_w^Hg_w\right|^4\})\nonumber\\
& =K^2\delta(\mathbb{E}\{\left| g_w^HN\phi_1^*\right|^4\}+q_t\mathbb{E}\{\left| g_w^Hg_w\right|^4\}+6q_t\sigma^2\mathbb{E}\{\left| g_w^Hg_w\right|^3\})\nonumber\\
&=K^2\delta M(q_t(M+1)(M+2)(6\sigma^2+(M+3)q_t\beta_w)+\dfrac{M(5M+11)\sigma^2\beta_w^2}{4}),
\end{align}
where 
$\delta=\sum_{i=1}^K \dfrac{1}{M^2(\eta p_t \beta_i+q_t \beta_w+\sigma^2)^2}$. 
$\mathbb{E}\{\left| g_w^Hg_w\right|^4\}$ 
and
$\mathbb{E}\{\left| g_w^Hg_w\right|^3\}$
are respectively third and fourth moment of a Chi-Square distribution. The Chi-square distribution moments with $k$ degree of freedom are given by 
\begin{align*}
\mathbb{E}\{\bm{X}^m\}=k(k+2)...(k+2m-2),
\end{align*}
where $k$ is the vector length. Since the eavesdropper channel vector is complex and the above equation is for a real vector, the degree of freedom for the eavesdropper channel is assumed to be $2M$.

\begin{align} \label{gwAB}
&2\mathbb{E}\left\lbrace g_w^HAg_wg_w^HB^Hg_w\right\rbrace \nonumber\\
&=2K\delta(\eta p_t\mathbb{E}\{g_w^HN\phi_1^*(N\phi_1^*)^Hg_w^Hg_w^H\sum_{i=1}^{K}g_ig_i^Hg_w\} \nonumber \\&+\eta p_tq_t\mathbb{E}\{g_w^Hg_wg_w^Hg_w^Hg_w^H\sum_{i=1}^{K}g_ig_i^Hg_w\})\nonumber \\&=2K\delta\eta p_tM(M+1)(\sigma^2+(M+2)q_t \beta_w)\beta_w^2\sum_{i=1}^{K} \beta_{i}^2
\end{align}
By substituting (\ref{gwAA}), (\ref{gwBB}) and (\ref{gwAB}) into (\ref{gw wi2}) the first term of (\ref{EZ2-E2Z}) is obtained as 
\begin{align}\label{gwA+Bgw2}
&\mathbb{E}\lbrace( \sum_{i=1}^{N}|\bm{g}_{w}^H \bm{w}_i|^2 )^2\rbrace\nonumber\\
&=2M \beta_ w^2 \sum_{i=1}^{K}\frac{\eta^ 2 p_t^2K(K+1) \beta_ i^2+2 \eta p_t \sigma ^2 \beta_ i}{M^2(\eta p_t \beta_ i+q_t \beta_ w+\sigma^ 2)^2} \nonumber \\
& + (M^2-M) \beta _w^2 \sum_{i=1}^K \frac{ \eta^ 2 p_t^2\beta_ i^2+2 \eta p_t \sigma^ 2 \beta_ i}{M^2(\eta p_t \beta_ i+q_t \beta_ w+\sigma^ 2)^2}\nonumber\\
& +\delta M(2M^2\sigma^4\beta_w^4+2\delta\eta p_tM(M+1)(\sigma^2+q_t \beta_w)\beta_w^2\nonumber\\
&+q_t\beta_w^3(M+2)(M+4)(6(\sigma^2+q_t\beta_w)+Mq_t\beta_w)).
\end{align}

\section*{Appendix C}
The partial derivatives of the secrecy rate lower bound according to $\theta$ and $\rho_k$ are detailed in (\ref{Rate secrecy diff rho}) and (\ref{Rate secrecy diff phi}). As it can be seen in (\ref{Rate secrecy diff rho}), the partial derivative according to $\rho_k$ is always positive and non-zero. Thus the secrecy rate lower bound is strictly increasing with $\rho_k$. 

\begin{align} \label{Rate secrecy diff rho}
 &\dfrac{\partial \,\text{R}_{\text{S,bound}}}{\partial \rho_k}=\dfrac{\sigma_s^2}{\rho_k\big(\rho_k\big(K\beta_k+(k-1)\beta_kW_E+\dfrac{\beta_k(q_t \beta_w+\sigma^2)}{\eta p_t\beta_k+q_t \beta_w+\sigma^2}\sigma_{ant}^2 \big)+\sigma_s^2\big)} \nonumber \\ \\
 &\dfrac{\partial \,\text{R}_{\text{S,bound}}}{\partial \theta}=\dfrac{\rho_k \eta p_t \beta_k^2 W_E(q_t \beta_w+\sigma^2)/(\eta p_t \beta_k+q_t \beta_w+\sigma^2)}{\big(\rho_k\big(K\beta_k+(k-1)\beta_kW_E+\dfrac{\beta_k(q_t \beta_w+\sigma^2)}{\eta p_t\beta_k+q_t \beta_w+\sigma^2}+\sigma_{ant}^2 \big)+\sigma_s^2\big)}\nonumber\\
 &+\dfrac{\big(\rho_k\big(K\beta_k+(k-1)\beta_kW_E+\dfrac{\beta_k(q_t \beta_w+\sigma^2)}{\eta p_t\beta_k+q_t \beta_w+\sigma^2}+\sigma_{ant}^2 \big)+\sigma_s^2\big)(q_t \beta_w+\sigma^2)}{\big(\rho_k\big(K\beta_k+(k-1)\beta_kW_E+\dfrac{\beta_k(q_t \beta_w+\sigma^2)}{\eta p_t\beta_k+q_t \beta_w+\sigma^2}+\sigma_{ant}^2 \big)+\sigma_s^2\big)(\eta p_t\beta_k+q_t \beta_w+\sigma^2)}\nonumber\\
 &+\dfrac{-\rho_{\text{Eve}}(W_E+1)\sum_{i=1}^{K}\dfrac{M q_t \dfrac{\eta p_t}{\theta} \beta_w^2 \beta_i}{(\eta p_t \beta_i+q_t \beta_w+\sigma^2)^2}+\rho_{\text{Eve}}\dfrac{Mq_t \dfrac{\eta p_t}{\theta}\beta_w^2\beta_k}{(\eta p_t \beta_k+q_t \beta_w+\sigma^2)^2}}{\rho_{\text{Eve}} \big((K\beta_w+\sum_{i=1}^K \dfrac{Mq_t \beta_w^2}{\sigma^2+q_t \beta_w+\eta p_t \beta_i })(W_E+1)-\dfrac{M q_t \beta_w^2}{\sigma^2+q_t \beta_w+\eta p_t \beta_k}+\sigma_{ant}^2\big)+\sigma_s^2}\nonumber\\
 &+\dfrac{\dfrac{\eta p_t}{\theta}\beta_k(\rho_{\text{Eve}} \big((K\beta_w+\sum_{i=1}^K \dfrac{Mq_t \beta_w^2}{\sigma^2+q_t \beta_w+\eta p_t \beta_i })(W_E+1)-\dfrac{M q_t \beta_w^2}{\sigma^2+q_t \beta_w+\eta p_t \beta_k}+\sigma_{ant}^2\big))}{\rho_{\text{Eve}} \big((K\beta_w+\sum_{i=1}^K \dfrac{Mq_t \beta_w^2}{\sigma^2+q_t \beta_w+\eta p_t \beta_i })(W_E+1)-\dfrac{M q_t \beta_w^2}{\sigma^2+q_t \beta_w+\eta p_t \beta_k}+\sigma_{ant}^2\big)+\sigma_s^2} \label{Rate secrecy diff phi}
\end{align}\\
\begin{figure}[!h]
	\centering
	%\hspace{0.1cm}
	\includegraphics[scale=0.4,trim=4.1cm 0.5cm 4cm 0.8cm,clip]{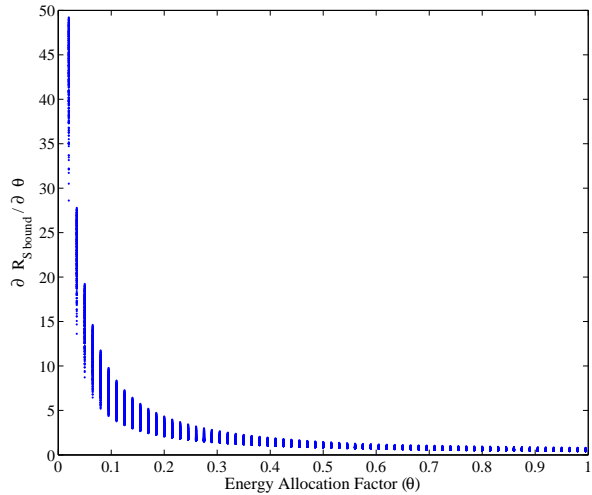}
	\caption{Positivity of derivative function of the secrecy rate lower bound according to harvested energy allocation factor ($\theta$). The derivative function of secrecy rate lower bound is plotted for various locations of the users and the active eavesdropper. $d_k \sim \mathcal{U}[10,20]$ (meters) and  $d_w \sim \mathcal{U}[10,20] $ (meters).}
	\label{fig:diffphipositive}
\end{figure}
Determining the positivity of the partial derivative function according to $\theta$ is not straightforward. Fig. \ref{fig:diffphipositive} shows the positivity of the partial derivative function according to $\theta$ for various distances of the users and  eavesdropper from the BS. As it can be seen, the partial derivative function is positive and non-zero. Therefore, the secrecy rate lower bound is strictly increasing with $\theta_k$.   
%\bibliographystyle{ieeetr}
%\bibliography{paper}
	
\end{document}